\documentclass[12pt, a4paper]{article}

\usepackage{amsmath}
\usepackage{amsfonts}
\usepackage{amssymb}
\usepackage{graphicx, rotating}
\usepackage{epsfig}
\usepackage{latexsym}
\usepackage{graphicx}
\usepackage{color}
\usepackage{amsmath,bm,amssymb}
\usepackage{cite}
\usepackage{slashed}
\usepackage{hyperref}
\hypersetup{colorlinks, citecolor=bluscuro, linkcolor=black, urlcolor=bluscuro}
\definecolor{rossos}{cmyk}{0,1,1,0.55}
\definecolor{bluscuro}{rgb}{0.15, 0.2, .85}
\definecolor{bluchiaro}{cmyk}{1,.3,0.,0.1}

\newcommand{\eq}[1]{Eq.~(\ref{#1})}

\newcommand{\pslash}{\!\not\! p}
\newcommand{\kslash}{\!\not\! k}

\newcommand{\be}{\begin{equation}}
\newcommand{\ee}{\end{equation}}
\newcommand{\bea}{\begin{eqnarray}}
\newcommand{\eea}{\end{eqnarray}}

\def\Tr{{\rm Tr\,}}

\begin{document}

\begin{titlepage}
\begin{flushright}
DESY 14-093
\end{flushright}
\vspace{.3in}

\vspace{1cm}
\begin{center}
{\Large\bf\color{black}
Taming 
Infrared Divergences\\[0.5cm]
in the Effective Potential}\\
\bigskip\color{black}
\vspace{1cm}{
{\large J.~Elias-Mir\'o$^{a,b}$, J.R.~Espinosa$^{a,c}$ and 
T.~Konstandin$^d$}
\vspace{0.3cm}
} \\[7mm]
{\it {$^a$\, IFAE, Universitat Aut{\`o}noma de Barcelona,
   08193~Bellaterra,~Barcelona}}\\
{\it {$^b$\, Dept.~de~F\'isica, Universitat Aut{\`o}noma de Barcelona, 08193~Bellaterra,~Barcelona}}\\
{\it $^c$ {ICREA, Instituci\'o Catalana de Recerca i Estudis Avan\c{c}ats, Barcelona, Spain}}\\
{\it $^d$ {DESY, Notkestr. 85, 22607 Hamburg, Germany}}
\end{center}
\bigskip

\vspace{.4cm}

\begin{abstract}
The Higgs effective potential in the Standard Model (SM), calculated perturbatively, generically suffers from
infrared (IR) divergences when the (field-dependent) tree-level mass of the Goldstone bosons goes to zero. 
Such divergences can affect both the potential and its first derivative and become worse with increasing loop order.
In this paper we show that these IR divergences are spurious, we perform a simple resummation of all 
IR-problematic terms known (up to three loops) and explain how to extend the resummation to cure all such
divergences to any order. The method is of general applicability and would work in scenarios other than the
SM. Our discussion has some bearing on a scenario recently proposed as
a mechanism for gauge mediation of scale breaking in the ultraviolet, in which it is claimed that the low-energy 
Higgs potential is non-standard. We argue that all non-decoupling effects from the heavy sector can be
absorbed in the renormalization of low-energy parameters leading to a SM-like effective theory.

\end{abstract}
\bigskip

\end{titlepage}

\section{Introduction} 

The effective potential is an extremely useful object that has played a central role
in solving and understanding numerous problems in particle physics and beyond. 
One of its recent uses, in the context of the Standard Model (SM), has been the study of
electroweak vacuum stability \cite{stab0,stab1,stab2,stab3}, a hot issue given the experimental values
of the top quark, $M_t=173.34\pm 0.76$ GeV \cite{mtop}, and the mass of the 
Higgs boson, $M_h=125.6\pm 0.57$ GeV \cite{mhiggs}, recently discovered by the LHC
\cite{higgsD}. For these masses the electroweak vacuum
is very close to the boundary in phase space that separates the region of stability up
to the Planck scale from the region of metastability, in which the electroweak vacuum is 
unstable.

In order to study this issue, a very precise calculation of the effective potential
at high field values was required. The state-of-the-art calculation, at NNLO \cite{stab0,stab1}, uses a
two-loop effective potential \cite{V2} with parameters running with 3-loop renormalization
group equations \cite{beta3} and matching between running and pole masses at
two-loop order \cite{stab0,stab1,stab2}. For the current values of the top and Higgs masses, 
absolute stability is disfavored and most likely the vacuum is metastable, although
its decay lifetime is enormously large compared with the age of the Universe.
Besides this stability issue, a precise knowledge of the Higgs effective potential 
in the SM is of interest at the electroweak scale itself as it offers a simple way of calculating
some subclasses of radiative corrections, as for instance an important part of those that enter 
in the matching relation between the Higgs quartic coupling and the pole Higgs mass \cite{stab0}.

Knowledge of the three-loop effective potential would therefore be of interest and
a first step in that direction was taken in Ref.~\cite{Martin3L}, that calculated the contributions
involving the top Yukawa and the strong gauge coupling to the three-loop potential 
in the SM (in Landau gauge and $\overline{\mathrm{MS}}$). This work encountered an 
infrared (IR) problem, previously noticed in Ref.~\cite{OtherIR}, in the contribution of Goldstone 
bosons to the potential, as described in Sec.~\ref{GoldCat}. One of the goals of this paper is to solve 
this IR problem. The solution, by means of a resummation of the relevant two-particle-reducible 
Goldstone contributions, is given in Sec.~\ref{Resum}. Such resummation takes 
care of an infinite series of IR divergences, that can be classified as the leading ones, and 
achieves an IR-finite 3-loop potential. Nevertheless, at even higher orders in the perturbative expansion,
there are further (sub-leading) IR divergences.  We devote Sec.~\ref{bey} to explaining the
origin of such subleading divergences and to discuss possible complications in their resummation. Then, in Sec.~\ref{AWOP} we 
show, using a Wilsonian-inspired organizing principle, how all these divergences can in principle 
be resummed to any desired order. In Sec.~\ref{HBC} we apply similar methods to a related type of IR problem that affects the potential 
(and its first derivative) when the (field-dependent) mass of the Higgs itself goes to zero.

These type of IR problems afflicting the effective potential are generic in quantum field theory. Although we study them 
in a particularly relevant and fairly generic theory, the SM, the resummation methods we find
are expected to be of more general applicability.​ In this regard, the solution of the IR problem 
that we present in Sec.~\ref{Resum} turns out to have a bearing on the 
scenario considered in Ref.~\cite{Abel}, in which scale invariance (assumed to be exact in the ultraviolet) 
is broken in a hidden sector and then communicated to the SM via gauge interactions. 
The low-energy Higgs potential obtained in Ref.~\cite{Abel} is of a peculiar form, with a mass term 
that has a logarithmic dependence on the Higgs field [$m^2\phi^2\log(\phi^2/\langle\phi\rangle^2)$]. 
In Sec.~\ref{LogPot}, we critically examine the derivation of this potential by \cite{Abel}, concluding that it suffers 
from IR artifacts. We argue that, after correcting for such IR problems, the Higgs potential obtained in such 
scenarios should be SM-like.  
We draw some conclusions in our final section and leave some technical details for an appendix.

\section{The Goldstone Boson Catastrophe} 
\label{GoldCat}

The SM effective potential $V(\phi)$ is calculated order by order in perturbation theory by
summing 1PI vacuum diagrams with Feynman rules derived (using Landau gauge in what follows) in a classical background
value $\phi$ for the Higgs field.  The Higgs scalar doublet is 
\be
\left(\begin{array}{c}
G^+\\
(h+\phi+iG^0)/\sqrt{2}
\end{array}
\right)\ ,
\ee
where $h$ is the Higgs (quantum) field and $G^0,G^+$ are the neutral and charged Goldstone bosons.
The field-dependent tree-level squared masses of the relevant SM fields are $T=h_t^2\phi^2/2$, for the top quark,
$W=g^2\phi^2/4$, for the $W^\pm$ gauge boson, $Z=g_Z^2\phi^2/4$, (where we use $g_Z^2=g^2+{g'}^2$), for the $Z^0$ gauge boson, $H=-m^2+3\lambda\phi^2$ for the Higgs and $G=-m^2+\lambda\phi^2$ for the Goldstones. In what follows, we neglect the masses of leptons and quarks other than the top. In the formulas above, $h_t$ is the top Yukawa coupling, $g$ and $g'$ are the 
$SU(2)_L$ and $U(1)_Y$ gauge couplings, respectively, and
the expressions given for the masses of Higgs and Goldstone bosons correspond to the tree-level potential 
\be
V^{(0)}(\phi) = -\frac{1}{2}m^2\phi^2+\frac{1}{4}\lambda\phi^4\ .
\ee

Using dimensional regularization and $\overline{\mathrm MS}$ scheme, the SM one-loop potential
in Landau gauge takes the well-known Coleman-Weinberg \cite{cw} form 
\bea
V^{(1)}(\phi) & = & -N_c\kappa T^2 (L_T-3/2)+\frac{3\kappa}{2} W^2(L_W-5/6)
+\frac{3\kappa}{4} Z^2(L_Z-5/6)\nonumber\\
&+&\frac{\kappa}{4} H^2(L_H-3/2)+\frac{3\kappa}{4} G^2(L_G-3/2)\ ,
\label{CW1}
\eea
where $\kappa=1/(16\pi^2)$, $N_c=3$ is the number of colors, and  $L_X=\log(X/Q^2)$, where $Q$ is the 
$\overline{\mathrm MS}$ renormalization scale. 
The two-loop SM potential was first computed in Ref.~\cite{V2}. A compact expression can be found, {\it e.g.}
in the appendix of Ref.~\cite{stab0}. The part of the three-loop potential obtained neglecting all couplings other 
than the top Yukawa and the strong gauge coupling has been calculated in Ref.~\cite{Martin3L}.

To identify the terms in the SM potential that cause IR problems in the limit $G\rightarrow 0$, we can perform
an expansion in powers of small $G$ of the known expressions for the potential. 
Generically, one expects to have in the $L$-loop potential contributions of order $G^n L_G^m$, where 
$n\geq 3-L$ and $m\leq L$, and we see that some of these contributions are IR-problematic. 
The most dangerous ones are the contributions to $V(\phi)$ that scale as $L_G^m$  (which appear at 3-loops and higher) 
or as inverse powers of $G$ (which appear at 4-loops and higher) as these give directly IR-divergent corrections to the 
potential (and all its derivatives) when $G\rightarrow 0$. Such pathological behavior cannot be accepted if the 
effective potential is to be of any use. 

Also dangerous are the terms in $V(\phi)$ that scale as $GL_G^m$, which are IR finite but cause a divergence in the first
derivative of the potential. This is problematic as finding the minimum of the potential requires solving the minimization equation $V'(\phi)=0$, which should be well defined for any $\phi$ value. In practice, this problem is often ignored, since higher loop contributions induce a small mass to the Goldstone bosons. This typically reduces the impact of the logarithmic Goldstone contributions in the minimum of the potential although this is not guaranteed in principle. Nevertheless, it is clear that a
solution should exist that eliminates this divergence, making the minimization of the potential a well-posed problem.\footnote{Along 
this line of argument, and working in the context of a finite temperature effective theory to study the SM electroweak phase transition, Ref.~\cite{Shapos} showed that similar IR divergences were artifacts of the expansion and would cancel out when computing a physical
quantity, like the value of the potential at the minimum (see appendix~A of Ref.~\cite{Shapos}).}

Less troublesome are contributions of order $G^2 L_G^m$ in $V(\phi)$, which are IR safe
but cause a divergence in $V''\equiv d^2V/d\phi^2$, as $V''$ would then contain terms of order $(G')^2L_G^m$,
divergent for $G\rightarrow 0$. This is potentially harmful for calculations of the Higgs mass, which is related precisely
to $V''$ evaluated at the minimum of the potential. The resolution of this problem is well-known (see {\it e.g.} appendix B of Ref.~\cite{mhos}): $V''$ reproduces the Higgs 2-point function at zero external momentum, while the Higgs pole mass 
is instead defined on-shell, with $p^2=M_h^2$. The calculation of the Higgs mass from the potential has to be corrected to take that into account and the correcting terms have precisely the same IR singularities as the potential term, canceling in the final result.
Notice that a similar solution from outside the potential itself cannot exist for the $GL_G^m$ terms as the minimization
equation should depend on the potential only. 

Explicitly, the troublesome terms in the SM potential are the following:
At one-loop, from $G$ loops without interactions, one has the $G^2 L_G$-terms
\be
V^{(1)}_{G^2 L_G} = \frac{3\kappa}{4}  G^2 L_G\ .
\ee
At two-loops one finds the following $G^2 L_G^m$-terms
\bea
V^{(2)}_{G^2 L_G} & = & \kappa^2 G^2 L_G\left\{\frac{1}{2}N_c h_t^2(3L_T-1)+
\frac{3}{8}g^2\left(-3-6 L_W+8W\frac{L_W-L_Z}{W-Z}\right)
\right.\nonumber\\
&+&\frac{3}{8}\frac{g^2}{W}H\left[3L_H-2L_W-L_Z-H\left(2\frac{L_H-L_W}{H-W}+\frac{L_H-L_Z}{H-Z}\right)\right]\nonumber\\
&+&\frac{3}{4}e^2\left(3L_Z+L_W-\frac{4}{3}\right) +\frac{3}{16}g_Z^2(5-6L_Z)\nonumber
\eea
\bea
&+&\left.\frac{3}{4} \lambda (3L_G-8)-\frac{e^2}{2}(3L_G-14)+\frac{3}{8}e^2(L_G-2L_W-5)
\right\}\ .
\label{v2g2lg}
\eea 
For later use, we have explicitly separated the three terms in the last row: the first comes from purely scalar loops, the second from the $G^\pm$-$G^\mp$-$\gamma$ two-loop diagram and the last one from the $W^\pm$-$G^\mp$-$\gamma$ contribution.  
In addition, there are also $G L_G^m$-terms:
\be
V^{(2)}_{G L_G} = \kappa^2 G L_G\left\{-3 N_c h_t^2T(L_T-1)   +\frac{9}{2}\lambda H(L_H-1) 
+\frac{3}{4}g^2W(3L_W-1)+\frac{3}{8}g_Z^2Z(3L_Z-1)\right\} .
\label{v2glg}
\ee
At three-loops, there are top-Yukawa terms of order $L_G$ given by~\cite{Martin3L}:
\be
V^{(3)}_{L_G} = 3\kappa^3 N_c^2 h_t^4 T^2(L_T-1)^2 L_G\ .
\label{v3lg}
\ee
There are other problematic 3-loop terms beyond this one, of order $\kappa^3 G L_G^m$ and $\kappa^3 G^2 L_G^m$,
(with $1\leq m\leq 3$) which are considered subleading in the analysis of Ref.~\cite{Martin3L} and not given there.

An indication that the IR Goldstone problem might be spurious is that it would be absent if instead of calculating the potential in
Landau gauge (which is the simplest to compute) one were to use unitary gauge, although that particular gauge has complications of its own. In any case, in the following section we show how to cure these IR problems in a very simple and direct way. Notice that this problem affects the behavior of the potential
at low-energy, near the electroweak vacuum. Goldstone contributions at very high values of the field are sizable also ($G\sim \lambda\phi^2$) and pose no particular threat for the potential. The possible impact of such IR divergences at large field values (those relevant for stability considerations) is therefore indirect: they could
affect the connection between potential parameters like the Higgs quartic coupling and observables like the Higgs pole mass.

\section{Resummation of IR Divergences}
\label{Resum}

The potentially dangerous Goldstone boson contributions to the effective potential  discussed in the previous section have a clear diagrammatic origin: at each loop level $L$
the most IR-divergent diagrams  involving Goldstone bosons consist of a ring of Goldstone bosons 
with $L-1$ insertions of one-loop Goldstone self-energies, as depicted
in Fig.~1.  We can draw an analogy with the resummation of the infinite series of one-loop diagrams with different number of 
background Higgs external legs that leads to the usual Coleman-Weinberg one-loop potential. Now
one would naively expect to be able to resum the infinite subset of Goldstone diagrams of the form given in fig.~1 into a one-loop resummed potential for Goldstone bosons of the form 
\be
\label{V1rp}
\delta_G V^{(1,r)} = \frac{1}{2} \int_p \log[-p^2 + G +\kappa\Pi_0(p^2) ] +
\int_p\log[-p^2 + G + \kappa\Pi_+(p^2) ]\ ,
\ee
where we use the short-hand notation
\be
\int_p \equiv \mu^{2\epsilon}\int \frac{d^Dp}{(2\pi)^Di}\ ,
\ee
with $D=4-2\epsilon$, $Q^2=4\pi e^{-\gamma_E}\mu^2$  and we have explicitly 
separated the neutral and charged Goldstone's contributions.
The explicit expressions for one-loop 2-point functions for neutral and charged Goldstone bosons are
given in the appendix.

\begin{figure}[t]
\begin{center}
\includegraphics[width=0.25\textwidth]{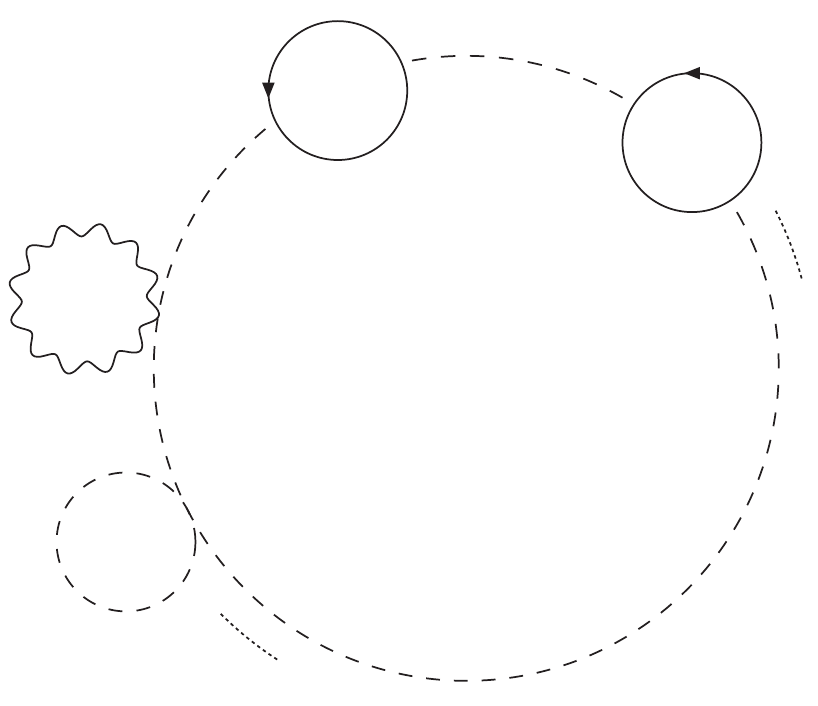} 
\parbox{14cm}{
\caption{\emph{Generic multiloop diagram with a Goldstone ring, dressed by Goldstone self-energies,
 that leads to IR problematic contributions to the Higgs potential.}\label{Gring}}
}
 \end{center}
\end{figure}

It is obviously impossible to perform the momentum integrals in the resummed potential
of Eq.~(\ref{V1rp}) with the full and complicated momentum dependence of the $\Pi_{0,+}(p^2)$
functions. However, in order to solve the IR problems we have described, this is not needed: it suffices
to use a low-momentum approximation for these two-point functions. This is appropriate as the problematic IR behavior
appears precisely when the momentum running in the Goldstone ring goes to zero, $p^2\sim G\rightarrow 0$. 
Let us write the low-momentum expansion of these two-point functions as 
\bea
\Pi_0(s) &= & \Pi_G + s \Pi'_0 +{\cal O}(s^2)\ ,\nonumber\\
\Pi_+(s) &= & \Pi_G + s \Pi'_+ +{\cal O}(s^2)\ ,
\label{piexp}
\eea
where $s=p^2$. The zero-momentum 2-point function $ \Pi_G$ is the one-loop correction
to the Goldstone bosons mass and must therefore be the same for charged and neutral Goldstone bosons
as both must vanish simultaneously at the (radiatively corrected) minimum of the one-loop potential.
An explicit expansion of the expressions for $\Pi_{0,+}(p^2)$ in the appendix confirms that result giving
\bea
\label{pig}
\kappa\Pi_G &=& 3 \kappa \lambda \left[G (L_G-1)+H(L_H-1)\right] -2N_c\kappa h_t^2 T(L_T-1)\nonumber\\
&+&\frac{3}{2}\kappa g^2 W (L_W-1/3)+\frac{3}{4}\kappa g_Z^2 Z (L_Z-1/3)\ .
\eea
This agrees with the result one can obtain directly from the one-loop Coleman-Weinberg potential (\ref{CW1})
simply as 
\be
\kappa\Pi_G = \frac{1}{\phi}\frac{d V^{(1)}}{d\phi}\ ,
\ee
which shifts the Goldstone boson masses ensuring they vanish in the minimum of the one-loop potential. 

If we insert this simple result for the two-point functions, that is $\Pi_{0,+}(p^2)\simeq \Pi_G$, in Eq.~(\ref{V1rp}), the momentum integral can be easily calculated
and gives
\be
V^{(1,r)}_{G}=\frac{3\kappa}{4}(G+\kappa\Pi_G)^2\left[\log\left(\frac{G+\kappa\Pi_G}{Q^2}\right)
-\frac{3}{2}\right]\ ,
\label{CW1R}
\ee
which is simply the usual Coleman-Weinberg correction from Goldstone bosons, with $G\rightarrow G + \Pi_G$. This shift amounts to
changing the tree-level Goldstone mass $G$ by its one-loop corrected version  $G + \Pi_G$.
Since the Goldstone bosons are massless in the minimum of the potential, $G$ and $\Pi_G$ are nominally 
of similar size close to the minimum of the potential. 
Upon expansion in powers of $\Pi_G$ 
\be
\delta_G V^{(1,r)} = \frac{3\kappa}{4}G^2(L_G-3/2) + \frac{3}{2}\kappa^2 G (L_G-1)\Pi_G+
 \frac{3}{4}\kappa^3 L_G \Pi_G^2 + \frac{1}{4G}\kappa^4\Pi_G^3 + {\cal O}(\kappa^5)\ ,
\ee
this simple resummed potential reproduces the most IR-problematic terms (coming from diagrams of
the type in fig.~1, with one-loop insertions) to all orders. Using the $\Pi_G$ expression written
above in (\ref{pig}), it can be checked that the two-loop terms of order $GL_G$ in (\ref{v2glg}) and the subset 
of 3-loop terms of order $L_G$ given in (\ref{v3lg}) are exactly reproduced by this expansion.  
The terms of order $G^2L_G$ in (\ref{v2g2lg}) are not exactly resummed but, as discussed before, do not pose a problem. 
The one-loop resummed term itself is harmless for the same reason. 

Notice also that $\Pi_G$ is IR safe,
so that the resummed potential (\ref{CW1R}) is IR finite when $G\rightarrow 0$. However, the first derivative
of the potential is still ill-defined, as it will involve a first derivative of the $GL_G$ term in $\Pi_G$. 
This pitfall can be easily avoided if one realizes that the Goldstone contribution to the self-energy in (\ref{pig})
is not really needed to cancel the IR divergent terms in (\ref{v2glg}) and (\ref{v3lg}). The reason is that the most 
dangerous ring diagrams of the type shown in Fig.~\ref{Gring} are those that do not involve $G$ contributions to the self-energy
insertions. 

We conclude that the resummed potential that cures the IR problems of the potential discussed before,
leading to IR finite $V(\phi)$ and $V'(\phi)$, is obtained as follows:
replace the usual one-loop Goldstone contributions by the term
\be
V^{(1,R)}_{G}=\frac{3\kappa}{4}(G+\kappa\Pi_g)^2\left[\log\left(\frac{G+\kappa\Pi_g}{Q^2}\right)
-\frac{3}{2}\right]\ ,
\label{CW1Rgood}
\ee 
where 
\be
\Pi_g = 3 \lambda H(L_H-1) -2N_c h_t^2 T(L_T-1)+\frac{3}{2} g^2 W (L_W-1/3)+\frac{3}{4} g_Z^2 Z (L_Z-1/3)\ ,
\label{pi}
\ee
and remove from higher order contributions in the usual effective potential the terms of two-loop and higher that arise 
from expanding (\ref{CW1Rgood}) in powers of $\kappa\Pi_g$. This resummed term (and its derivative) are finite both
for $G\rightarrow 0$ and $G+\kappa\Pi_g\rightarrow 0$.

Before proceeding, let us examine more closely what we have done so far: First, notice that the resummation 
of harmful diagrams seems to be straightforward, but actually it is not. In particular, it is well known from similar 
resummations in other contexts ({\it e.g.} Daisy resummation of diagrams in the finite temperature effective potential) that the two-loop contributions are not well reproduced in general. The simplest
example of this can be illustrated by a simple $\lambda \phi^4$ model,  that describes a self-interacting boson $\phi$
with squared mass $m^2>0$. The one-loop self-energy, which is independent of the external momentum, is  
$\kappa \Pi = 6 \lambda\kappa m^2 \, \left(L_{m^2} - 1 \right)$
and the resummed expression (\ref{V1rp})  gives a two-loop contribution 
%
%
\be
\frac{1}{2}\int_p\, \frac{\kappa\Pi}{m^2-p^2 } \, ,
\ee
that should be compared with the true contribution coming from a eight-shaped diagram, which is smaller by a factor 2. 
This mismatch in the symmetry factors results from the fact that in such two-loop diagram it is ambiguous to determine which propagator belongs to the self-energy and which to the external ring. 

The same issue arises in our case for two-loop diagrams that have two Goldstone propagators,
as in the example shown in Fig.~2. For these diagrams it is in principle ambiguous to decide which of the two Goldstone propagators belongs in the external 
Goldstone ring and which in the one-loop self-energy dressing the ring (no such ambiguity appears for diagrams 
of higher order of the type in Fig.~1). Naively, the symmetry factor for such two-loop diagrams is a factor 2 smaller
than needed to match the result from the expansion of the one-loop corrected potential. 
However, one should notice that IR-divergent $L_G$ terms arise in loop integrals 
when the momentum in the Goldstone propagator is of order G.
\begin{figure}[t]
\begin{center}
\includegraphics[width=0.25\textwidth]{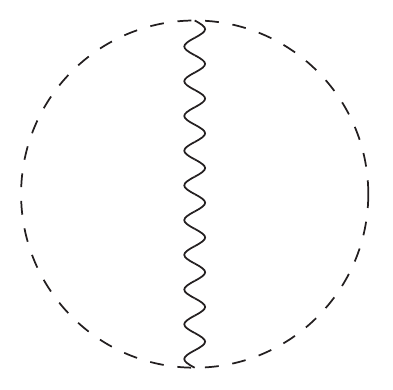} 
\parbox{14cm}{
\caption{\emph{Two-loop diagram with two Goldstones  and a $Z$ that naively would lead to the wrong combinatorics for resummation of IR divergences.}}
}
 \end{center}
\end{figure}
Then it can be shown that the $GL_G$ two-loop terms come from regions in the momentum integral for which only one
of the Goldstone propagators is soft, while the other carries large momentum.\footnote{A direct way of checking this is to use the method of regions \cite{regions} to obtain an expansion of the momentum integrals in powers of small masses. This method allows one to identify precisely the momentum range from which IR-problematic terms come from. For instance, the loop integral $I(G,G,Z)$ that appears when evaluating the diagram in Fig.~2, contains a term $\sim \kappa^2 GL_G$ coming from loop momenta $k_1^2\sim G$ and  $k^2_2\sim Z$, where $k_{1,2}$ are the momenta carried by each Goldstone line.}
This unambiguously selects which Goldstone should be counted as belonging in the ring and which one in the 2-point function. In other words, for a two-loop diagram with two Goldstone propagators this gives an extra factor of 2 for the soft-Goldstone/hard-Goldstone contribution. On the other hand,  when both Goldstone momenta are soft one obtains $G^2 L_G^2$ terms, which are not dangerous and in principle need no 
resummation. In fact, we have already seen that the terms of order $G L_G$ are reproduced correctly by the resummed expression, 
while the terms of order $G^2 L_G$ are not.

\section{Beyond Leading IR Divergences\label{bey}}

In the previous section we have shown that all the IR divergent terms in the SM potential calculated so far
(that is, terms that cause $V$ or $V'$ to diverge in the limit of zero tree-level mass for the Goldstones, $G\rightarrow 0$) 
can be resummed. The simple recipe is to shift $G$ in $V^{(1)}$ by its zero momentum one-loop two-point function (including in it only contributions
from particles with nonzero masses in the $G\rightarrow 0$ limit). Next we show that the leading  IR divergences\footnote{By \emph{leading} we mean those that diverge at the highest rate at a fixed loop order, see below.} as $G\rightarrow 0$ are resummed 
in the same way at arbitrary order in the perturbative expansion; along the way we will learn more about the IR divergent structure of the effective potential.

\begin{figure}[t]
\begin{center}
\includegraphics[width=0.25\textwidth]{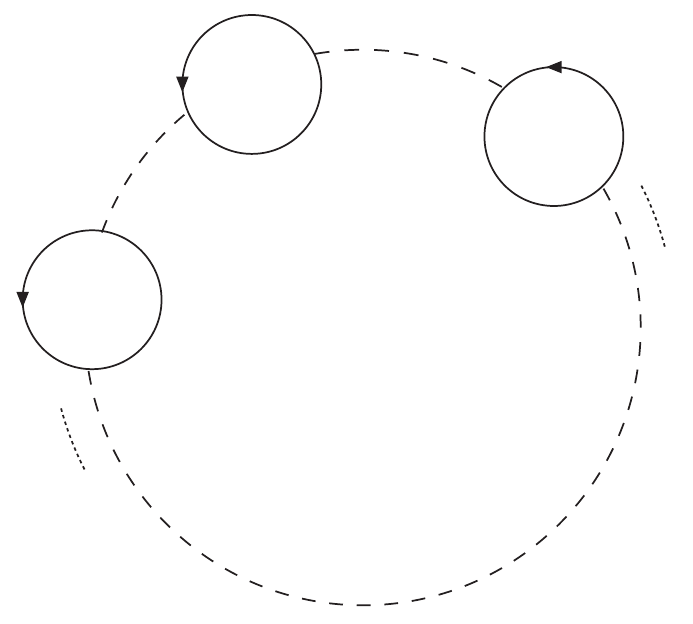}
\end{center}
\begin{center}
\parbox{14cm}{
\caption{
\label{fig:RL}
\emph{$L$-loop diagram with a neutral Goldstone ring adorned by $(L-1)$-iterated top-quark loops.
}}}
\end{center}
\end{figure}

To begin with, consider the contribution to the effective potential from the $L$-loop diagram with a neutral Goldstone ring adorned by $(L-1)$-iterated top-quark loops shown in Fig.~\ref{fig:RL}. It gives\footnote{We follow the Feynman rules and conventions of Ref.~\cite{Aoki}.}
\bea
   R_L&=& \frac{-1}{2(L-1)} \int_p \left(\frac{1}{G-p^2}\right)^{L-1} 
\left\{\frac{-N_c y_t^2}{2}\int_k \text{Tr}\left[\frac{\pslash+\kslash+m_t}{m_t^2-(p+k)^2}(i\gamma_5)\frac{\kslash + m_t}{m_t^2-k^2}(i\gamma_5)\right]\right\}^{L-1}  \nonumber \\[0.2cm]
   &=&  \frac{-1}{2(L-1)} \int_p \left(\frac{1}{G-p^2}\right)^{L-1}\left[-\kappa\Pi_1(p^2)\right]^{L-1} \, ,   \label{exact37}
\eea
where $\Pi_{1}(p^2)$ is the top contribution to the one-loop Goldstone's two-point function.  In the momentum region $p^2\sim G$ (the one responsible for the IR divergences) and $k^2\sim m_t^2=T$ the integrand of \eq{exact37} can be approximated by a low-momentum expansion, $\kappa\Pi_1(p^2)=\kappa\Pi_1(0) + {\cal O}(p^2)$. Notice that extra powers of ${\cal O}(p^2)$ generate extra powers of $G$ upon integration and hence less divergent contributions as $G\rightarrow 0 $.  One gets 
  \be
 R_L=
-\frac{1}{2(L-1)}  \int_p  \left(\frac{1}{G-p^2}\right)^{L-1} \left[- \kappa\Pi_1(0)\right]^{L-1}  [1+ {\cal O}(G)]\label{mostIR} \, .  
\ee
This result exactly agrees\footnote{In following this method of regions \cite{regions} we integrate
an expression, expanded assuming small momentum,  over the whole domain of loop momentum 
 (i.e. outside the regime of validity of the expansion). This procedure introduces spurious UV divergences,
but they cancel between the different momentum regions  in which one should expand the integrand \cite{regions}
to calculate its $G$-expansion.}  with the Goldstone contribution to the L-loop term of the resummed potential:  
\be
 \frac{1}{2}\int_p\log \left[-p^2 + G + \kappa \Pi_1(0) \right] =
\frac{1}{2}\int_p \log \left[-p^2 + G \right] -\sum_{L=2}^\infty
  \frac{1}{2(L-1)}\int_p\left[\frac{-\kappa\Pi_1(0)}{G-p^2}\right]^{L-1} \, . \label{resum}
\ee
The result in \eq{mostIR} gives  the leading IR divergent contribution
 from top loops at order $\kappa^L (y_t^2)^{(L-1)}$. By performing the momentum integral, \eq{mostIR} gives terms
$\propto G (\log G-1 )$ for $L=2$, $\log G$ for $L = 3$, and $\propto G^{3-L}$, for $L> 3$. Notice that this
method reproduces in a straightforward way the $L=3$ top contribution obtained by Ref.~\cite{Martin3L}:
$3 R_3=[3 \kappa^3\Pi_1(0)^3/4] \log G = 3 \kappa^2  y_t^4N_c^2 L_G T \left(L_T-1\right)$,
with the extra factor $3$ accounting for the Goldstone multiplicity. 

For $L>3$, the ${\cal O}(p^2)$ terms neglected in going from \eq{exact37} to \eq{mostIR}, also lead to IR divergences in $V$ or $V^{\prime}$. These divergences blow up at a lower pace than the leading ones we resum in \eq{resum} by shifting the Goldstone mass by its one loop correction $\kappa\Pi_1(0)$, but of course they are as pathological as the leading ones. 
We can go one step further in the resummation by including also the effect of $\Pi'_{0,+}$ in the 
approximation to the Goldstone two-point functions (\ref{piexp}). The momentum integral in (\ref{V1rp}) can still be performed and leads to
\bea
V^{(1,r')}_{G}&=&\frac{\kappa}{4}\left(\frac{G+\kappa\Pi_g}{1-\kappa\Pi'_0}\right)^2\left\{\log\left[\frac{1}{Q^2}\left(\frac{G+\kappa\Pi_g}{1-\kappa\Pi'_0}\right)\right]-\frac{3}{2}\right\}\nonumber\\
&+&\frac{\kappa}{2}\left(\frac{G+\kappa\Pi_g}{1-\kappa\Pi'_+}\right)^2\left\{\log\left[\frac{1}{Q^2}\left(\frac{G+\kappa\Pi_g}{1-\kappa\Pi'_+}\right)\right]-\frac{3}{2}\right\}\ ,
\label{V1ri}
\eea
where $\Pi_g$, given in (\ref{pi}), does not include the loops of $G$'s. The top contributions to $\Pi'_{0,+}$ are given in the Appendix,
see Eqs.~(\ref{Pip0}) and (\ref{Pipplus}).  One can check by expanding (\ref{V1ri})
\bea
V^{(1,r')}_{G}&=& \frac{\kappa}{4}G^2L_G+\frac{\kappa^2}{2}\left(\Pi_g GL_G+\Pi'_0G^2L_G\right)
+\frac{\kappa^3}{4}\left[\Pi_g^2L_G+4\Pi_g\Pi'_0 GL_G+3\left(\Pi'_0\right)^2G^2L_G\right]\nonumber\\
&+&\frac{\kappa^4}{4}\left[\frac{\Pi_g^3}{3G}+2\Pi_g^2\Pi'_0L_G+6\Pi_g\left(\Pi'_0\right)^2GL_G+
4\left(\Pi'_0\right)^3G^2L_G\right] \nonumber\\
&+& (\mathrm{ IR}\, \mathrm{finite}\, \mathrm{ terms})+ {\cal O }(\kappa^5) +2\left\{\Pi'_0\rightarrow \Pi'_+\right\}
\label{V1riexp}\eea
that the top two-loop terms of order $G^2L_G$ given in (\ref{v2g2lg}) are also reproduced and the same expansion will capture top IR divergent terms at higher order. 

Table~\ref{tabDiv} shows the hierarchical structure of the IR divergences we have discussed so far, showing the approximation in $\Pi(p^2)$ corresponding to a given divergence at each loop level. We can also see from the discussion above that we can perform an alternative expansion of the resummed potential (\ref{V1ri}), or a reordering of the previous expansion
(\ref{V1riexp}), if we keep $\tilde G\equiv G+\kappa \Pi_g$ unexpanded, as
\be
V^{(1,r')}_{G}=\tilde G^2 L_{\tilde G} \left[\frac{\kappa}{4}+\frac{\kappa^2}{2}\Pi'_0+\frac{3\kappa^3}{4}\left(\Pi'_0\right)^2+
\kappa^4\left(\Pi'_0\right)^3+{\cal O}(\kappa^5)\right]+
 (\mathrm{ IR}\, \mathrm{finite}\, \mathrm{ terms})+2\left\{\Pi'_0\rightarrow \Pi'_+\right\}.
\label{Vrbetter}
\ee
This teaches us that reorganizing the perturbative expansion around the shifted $\tilde G$ is the most efficient way of
dealing with the subleading IR divergences, that will require no special treatment. This can be shown to hold also for higher
orders in the momentum expansion of $\Pi_1(p^2)$ directly at the level of the momentum integral of the corrected 
propagator:
\be
 \frac{1}{2}\int_p \log \left[-p^2 + G + \kappa \Pi_1(p^2) \right] =
\frac{1}{2}\int_p\log \left[-p^2 + \tilde G \right]
 -\sum_{L=2}^\infty
  \frac{1}{2(L-1)}\int_p\left[\frac{-\kappa\, \delta\Pi_1(p^2)}{\tilde G-p^2}\right]^{L-1} ,  \label{resum2}
\ee
where $\delta \Pi_1(p^2)\equiv \Pi_1(p^2)-\Pi_1(0)$ starts at ${\cal O}(p^2)$, making the IR behavior of the momentum integrals in the sum never worse than $\tilde G^2 L_{\tilde G}$.

From this we can derive a simple recipe to implement the resummation: identify the $G^{2+n}L_G^m$ contributions
in the potential and shift in them $G\rightarrow \tilde G=G +\kappa \Pi_g$, correcting for this by removing all higher order terms in the expansion
of such shifted contributions in powers of $\kappa \Pi_g$. In the next section we will generalize and justify further this recipe.

{\renewcommand{\arraystretch}{1.6} 
\begin{table}[t]
\begin{center}
    \begin{tabular}{| p{2cm} || p{2cm} |  p{2cm}  | p{2cm}  p{0.3cm} }     
               \hline & \centering$\Pi_1(0)$  & \centering$\Pi_1^\prime(0)$  &   \centering$\Pi_1^{\prime\prime}(0)$  & $\cdots$ \\ \hline\hline
            \centering $ R_1$  & \centering    $G^2 \log G$  & & &   \\ \hline
            \centering  $R_2$ &  \centering $G \log G $ &  \centering  $G^2 \log G$ &  &  \\ \hline
         \centering     $R_3$&  \centering $ \log G$  &  \centering  $G \log G $ &  \centering $G^2 \log G$  &  $\cdots$\\ \hline
        \centering      $R_4$&  \centering $ 1/G$  &   \centering $ \log G$&   \centering $G \log G $  &  $\cdots$ \\ \hline
       \centering       $R_5$&   \centering  $ 1/G^2$  &  \centering  $ 1/G$& \centering $ \log G$  & $\cdots$  \\  
       \centering       $\vdots$&\centering     $\vdots$  &\centering  $\vdots$ &   \centering $\vdots$ &  $\ddots$   \\ 
    \end{tabular}
    \end{center}
    \caption{A class of IR divergences (for $G\rightarrow0$) in the $L$-loop contribution to the effective potential from the $L$-loop Feynman diagram $R_L$ with a Goldstone ring dressed with $L-1$ insertions of the one-loop Goldstone's two-point function. } \label{tabDiv}
\end{table}
}

Nevertheless, Table~\ref{tabDiv} does not exhaust all the possible IR divergent terms that can appear in the high-loop effective 
potential contributions. First, there are other diagrams involving
other particles besides tops and second, one can have diagrams with more complicated topologies. For instance, still at loop-level $L$,  instead of $L-1$ one-loop insertions in a single Goldstone ring we can have $L-3$ one-loop insertions and a two-loop self-energy insertion and so on. This will in general reduce the degree of the associated IR divergence. In any case, an extension of the procedure just described is possible, as we discuss in the next section.

Concerning the contributions from loops of other particles, one can encounter a few complications in comparison with the
top case just discussed:
\begin{itemize}
\item Consider first two-loop contributions to the potential from purely scalar diagrams. It can be checked that
the contributions of order $G^2L_G$ coming from the setting-sun diagram $G$-$G$-$H$ are not correctly resummed by 
including $\Pi'_{+,0}$ terms as for the top loops. 
\item We already  saw in the previous section that corrections from the eight-shaped diagram
$G$-$G$ were not resummed by $\Pi_{+,0}(0)$. 
\item In the case of corrections involving gauge bosons we find a different problem:
in resummed expressions like (\ref{Vrbetter}), we should make sure that $\Pi'_0$ and $\Pi'_+$ are also IR safe.
However, as can be seen in Eq.~(\ref{Pipplus}) of the Appendix, $\Pi'_+$ receives contributions from a $G^+$-$\gamma$ loop (next-to-last term) that includes an IR divergent term ($\sim L_G$)  and a $W^+$-$\gamma$ loop (last term) that includes a $\log(-s/Q^2)$ term.
\end{itemize}

The latter contribution requires special treatment as it cannot be simply substituted in the new resummed one-loop potential (\ref{V1ri}) or (\ref{Vrbetter}), which assume $\Pi'_+$ does not depend on the external momentum. However, it is straightforward to take care of this complication, at least in a perturbative expansion, simply computing the momentum integrals with additional factors of $\log(-p^2/Q^2)$ in the integrand. Writing 
\be
\Pi(s)=\Pi(0)+s\ \Pi'(0)+s\ \delta\Pi'(0)\log(-s/Q^2)+{\cal O}(s^2)\ ,
\ee
one finds
\be
 \frac{1}{2}\int_p \log \left[-p^2 + G + \kappa \Pi(p^2) \right] =\tilde G^2L_{\tilde G}\left\{\frac{\kappa}{4}+\frac{\kappa^2}{2}\left[\Pi'(0)+\frac{1}{2}\delta\Pi'(0)\right]+{\cal O}(\kappa^3)\right\}\ ,
\ee
with $\tilde G\equiv G +\Pi(0)$. Using such result one can then show that the IR-divergent two-loop term from the $W^+$-$G^-$-$\gamma$ diagram [the last term in \eq{v2g2lg}] is then exactly resummed too.

All the problematic contributions just discussed have in common that they involve light degrees of freedom (Goldstones and/or photons). In the next section we show how to deal with such complications.

\section{Resumming all IR Divergences to Arbitrary Order\label{AWOP}}

\subsection{A Wilsonian Organizing Principle}

There is a powerful organizing principle that we can use to guide the resummation of IR-problematic contributions to the potential
and to justify previous choices, like using $\Pi_g$ in \eq{pi}
instead of $\Pi_G$ in \eq{pig}, or dropping some parts of $\Pi'_+$. It relies on using a Wilsonian effective theory approach (implemented through the method of regions \cite{regions}) to separate the contributions
to the effective potential from degrees of freedom that are light or heavy in the limit $G\rightarrow 0$. The only light degrees of freedom are the neutral and charged Goldstones, the photon and the gluon. All the rest ($W,Z,H,T$ and even other lighter fermions) can be considered heavy. In addition, we will treat hard Goldstones (with momentum $k^2\gg G$) as heavy, while only
soft Goldstones (with $k^2\sim G$) are considered as light. In what follows, we use `light' and `heavy' in this extended sense.\footnote{We use the term \emph{Wilsonian} in the sense that the method relies on identifying the relevant degrees of freedom that contribute in the region of low  momenta. However, we are not proposing to compute the potential in the effective field theory with heavy particles integrated out but rather to identify the diagrammatic origin of the various IR-problematic terms to perform the resummation of the higher loop diagrams, see below.}

There are contributions to the potential that only involve heavy fields, which can be considered separately and do not suffer from IR divergences. Contributions to the potential that only involve light states are also IR safe: by dimensional analysis they can only give contributions proportional to $G^2L_G^m$, as $G$ is the only mass scale for them. The troublesome diagrams involve heavy and light fields. The presence of the heavy mass scale allows now IR problematic contributions to the potential like $XGL_G$ or $X^2L_G$ or
$X^3/G$, etc., where $X=T,W,Z,...$, as we have seen before. The key point is then to extract from those mixed diagrams the IR troublesome parts, which can be unambiguously identified as coming from diagrams that only involve light fields, with heavy particles
integrated out. In practice, this integrating out results from expanding propagators of heavy particles in the regions of low momenta leading to IR divergences. 

Let us use the $H$-$G^0$-$G^0$ two-loop contribution to the potential as an illustrative example. This contribution is proportional to 
the following mixed momentum integral
  \bea 
\begin{minipage}[h]{0.16\linewidth}
        \vspace{0pt}
        \includegraphics[width=\linewidth]{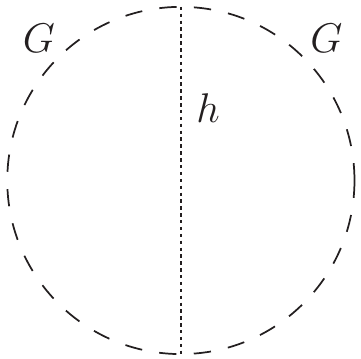}
   \end{minipage}\propto   I(G,G,H) =
\int_p\int_q\frac{1}{(G-k^2)}\frac{1}{(G-q^2)}\frac{1}{[H-(k+q)^2]}  \, . \nonumber
\label{IGGH}\eea 
The function $I(G,G,H)$ is known analytically (see {\it e.g.} Ref.~\cite{V2,V2M}) and the IR-problematic terms when $G\rightarrow 0$ can be easily obtained in an expansion in $G/H$ (the expansions in the Appendix of Ref.~\cite{EZ} can be useful in such derivation). One gets
\be
I(G,G,H) =-2\kappa^2  G  L_G (L_H-1) + \kappa^2\frac{G^2}{H} \left[L_G^2+L_G(1-2L_H)\right]  +\text{finite terms} \ ,
\label{exact26}
\ee
where `finite terms' means terms that give finite $\partial_G I$ or $\partial_G^2 I$ when $G\rightarrow 0$.

We will now obtain the singular terms in \eq{exact26} directly by using the method of regions. Considering that each Goldstone propagator 
can be hard or soft, we can split the momentum integral as
\be
I(G,G,H)=I(G_s,G_s,H)+2I(G_s,G_h,H)+I(G_h,G_h,H)\ ,
\ee
where $G_s$ ($G_h$) means a soft (hard) Goldstone is considered, corresponding to the different regions of momenta  in the integrand of \eq{exact26}. Let us discuss each case separately. 

\begin{itemize}
\item {\bf $ \bf I(G_s,G_s,H)\ (\text{with}\ k^2,q^2\sim G)$}:
In this region of momenta we can `integrate out' the Higgs by Taylor expanding in powers of $1/H$ the Higgs propagator in \eq{exact26}, obtaining
\be
I(G_s,G_s,H)=\frac{1}{H}\left[\int_q\frac{1}{(G-q^2)}\right]^2
+ {\cal O}\left(G^3/H^2\right)=
\kappa^2 \frac{G^2}{H} \left(L_G -1\right)^2+ {\cal O}\left(G^3/H^2\right)   \, .
\label{ss}
\ee

\item {\bf $\bf I(G_s,G_h,H)\ (\text{with}\ k^2\sim G,\ q^2 \sim H)$}:
In this momentum region we can expand the hard Goldstone propagator in powers of $G/q^2$ and the Higgs propagator
in powers of $k^2/(H-q^2)$. After performing the corresponding integrals, one gets
\bea
\label{wtdf}
2I(G_s,G_h,H)&=& 2 \kappa^2 G \left(L_G-1\right) \left[1-L_H +\frac{G}{2H}\left(3-2L_H\right)\right]+ {\cal O}\left(G^3/H^2\right) \ .
\label{sh}
\eea

\item {\bf $\bf  \ I(G_h,G_h,H)\ (\text{with}\ k^2, q^2 \sim H)$}:
In this case one can expand the two Goldstone propagators in  \eq{exact26} in powers of $G/q^2$, $G/k^2$. One gets
\be
I(G_h,G_h,H)= \int_p\int_q\frac{1}{k^2q^2[H-(k+q)^2]} \left[1+{\cal O}(G)\right] \, ,
\ee
which contains no terms $\propto L_G$. 
\end{itemize}

\begin{figure}[t]
\begin{center}
\includegraphics[width=0.7\textwidth]{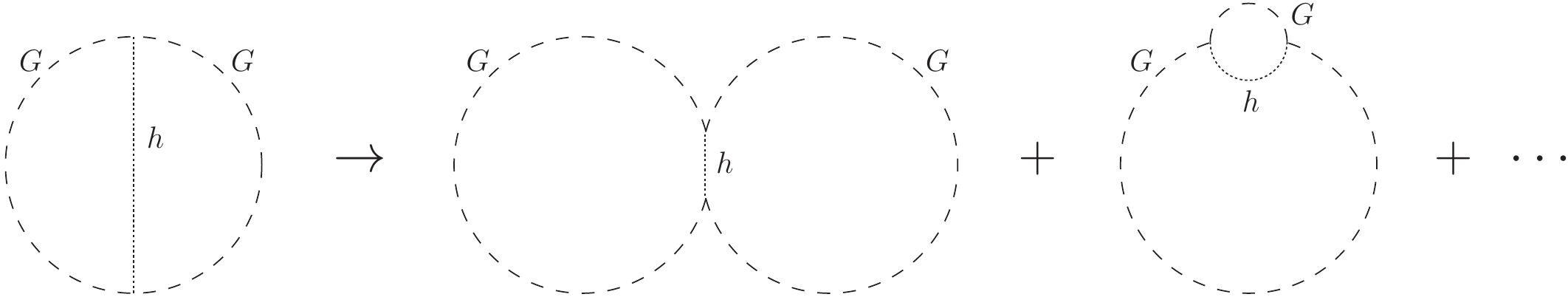}
\end{center}
\begin{center}
\parbox{14cm}{
\caption{
\label{fig:IGGH}
\emph{Diagrammatic representation of the decomposition of the two-loop momentum integral $I(G,G,H)$ from (\ref{IGGH})
(left diagram) in contributions with two soft Goldstones (first diagram on the right), one soft Goldstone (second diagram on the right)
plus contributions with hard Goldstones only (ellipsis). 
}}}
\end{center}
\end{figure}

We  see that the IR divergent terms in the sum of Eqs.~(\ref{ss}) and (\ref{sh}) exactly reproduce those in the exact result,
\eq{exact26}. There is a direct diagrammatic interpretation of $I(G_s,G_s,H)$ and $I(G_s,G_h,H)$ shown in Fig.~\ref{fig:IGGH}.
The first diagram in the right corresponds to $I(G_s,G_s,H)$ and shows the Higgs propagator reduced in size to indicate it has been 
integrated out. The second diagram on the right corresponds to $I(G_s,G_h,H)$, with $G_h$ and $H$ giving a contribution to the polarization of the soft Goldstone.

This splitting of $I(G,G,H)$ is crucial to order the resummation to be performed for a whole series of diagrams that radiatively correct
this diagram. Consider first the $I(G_s,G_h,H)$ part: it can be considered as the lowest term in a tower of corrections that can be
resummed in the same way we did in Sec.~\ref{Resum}, with a single Goldstone propagator and a self-energy insertion. The $GL_G$ term in \eq{wtdf} corresponds in this particular case to the $G-H$ contribution to $\Pi_g$ in Eq.~(\ref{pi}), which was already correctly resummed. The $G^2L_G$ term in \eq{wtdf} corresponds to a $\Pi'(0)$ contribution to the same kind of resummed diagram. 

However, we also see that $I(G_s,G_s,H)$ should be considered as the starting point of a different resummation: a eight-shaped diagram with soft Goldstones with resummed propagators, leading to a contribution proportional to $(G+\Pi_g)^2(L_{G+\Pi_g}-1)^2$.
We then understand that the reason for the failure of the resummation of  the $\lambda G^2L_G$ terms in \eq{v2g2lg} discussed at the end of the previous section is due to the fact that a part of these $\lambda G^2L_G$ terms [those coming from $I(G_s,G_s,H)$] has to be included in this second resummed diagram. 

In a similar way one is able to resolve the complications discussed at the end of the previous section. Consider
the $G^+$-$W^-$-$\gamma$ mixed diagram: the same procedure of decomposing the full diagram into diagrams with only light degrees of freedom will assign the IR-problematic terms to a diagram with one single soft Goldstone with a 2-point function
insertion (from a $W^-$-$\gamma$  loop), that is $I(W,G,0)=I(W,G_s,0)+$ finite terms.

A similar attempt at splitting the $G^+$-$G^-$-$\gamma$ two-loop contribution to the potential shows that the IR divergent terms
come from both Goldstones being soft because the mixed term $I(G_s,G_h,0)$ evaluates to zero. In this case, the full diagram
has light fields and no splitting is needed: that diagram has to be considered separately.

We can extend this procedure to any mixed diagram with light and heavy degrees of freedom, reducing any such diagram to a sum of 
diagrams involving only light fields. Such diagrams can then be considered either as contributions to some resummed term of lower order in the perturbative expansion or as new starting points for resumming full towers of corrections to the potential. The classification of different contributions is unambiguous and no overcounting problems arise. The key effect on light fields of integrating out the heavy degrees of freedom following this procedure is to give the soft Goldstones a mass shift $\Pi_g$ [which is given at one-loop by (\ref{pi}) but will
have contributions at higher loop orders from more complicated diagrams] and to leave behind diagrams that only involve propagators of light fields, so that their contribution to the potential is always proportional to $(G+\Pi_g)^2$ or higher powers [like $(G+\Pi_g)^{2+n}/T^n$]. As a result, IR divergences from Goldstones when $G\rightarrow 0$ or $G+\Pi_g \rightarrow 0$ are absent.

This justifies the use of the resummation recipe already discussed in a more restricted setting in the previous section.
We can state this recipe as follows: if the potential, expanded in small $G$, contains terms of the form $\delta V$ shown below on the right (with $n,m\geq 0$), add to the potential the zero piece ($\Delta V=0$)
in the left
\be
\delta V = \alpha\ \frac{G^{2+n}}{X^n}L_G^m \quad \Rightarrow \quad \Delta V=  \alpha\ \frac{\tilde G^{2+n}}{X^n}L_{\tilde G}^m
-\alpha\frac{1}{X^n} \sum (G+\kappa \Pi_g)^{2+n}L_{G+\kappa \Pi_g}^m\ ,
\label{recipe}
\ee
where $\alpha$ is a combination of couplings, $X$ is a mass scale
like $T, W, Z$ and the last term in $\Delta V$ is a series expansion in powers of $\kappa \Pi_g$. This series 
cancels out the divergent terms associated with the term $\delta V$ leaving behind the first (resummed) term in $\Delta V$, which 
is IR safe.
Depending on the order at which the resummation is needed, one might need to evaluate $\Pi_g$ beyond one loop. In that case, only
corrections from heavy fields (including hard Goldstones, with large momentum, $k^2\gg G$) should enter in that calculation.

\subsection{2PI Effective Action Approach}
Alternatively, a clean way of achieving the correct resummation at all orders 
is by using the {\it two-particle irreducible} (2PI) effective action~\cite{Cornwall:1974vz}. 
In the 2PI scheme, the effective action reads
\be
\label{eq:2PIaction}
\Gamma(P,\phi) = I(\phi) + \frac{i}2 \hbar \, \Tr \log P^{-1} \,
+ \frac{i}2 \hbar \, \Tr \, (p^2 - G) P \,
+ \Gamma_2(P, \phi) \, ,
\ee
where $I(\phi)$ denotes the tree-level action, $P$ the full propagator and $\Gamma_2$ contains only
{\it two-particle irreducible} diagrams. Broadly speaking, this result can  be explained as follows:
suppose one adds and subtracts an extra term in the Goldstone quadratic part of the action, so that
there is an added term in the Goldstone propagator, while the subtracted term is treated as an ${\cal O}(\hbar)$ counter-term,
to be inserted in perturbative calculations. If the added term matches the full (1PI) self-energy $\Pi(p^2)$, the counter-term cancels all diagrams where a subdiagram contributes to the self-energy. These are by definition the two-particle reducible
diagrams. However, there is a mismatch for certain diagrams (for example the eight-shaped diagram we have discussed above). This mismatch is taken care of by the next-to-last term in (\ref{eq:2PIaction}), see Ref.~\cite{Blaizot:2003an}. 
By construction, this scheme avoids IR issues that arise from
introducing a large number of self-energy insertions.
The non-perturbative proof is provided in Ref.~\cite{Cornwall:1974vz}.

The full propagator is in this scheme obtained by using the functional relation
\be
\frac{\partial \Gamma_2}{\partial P} = - \frac{i}2 \hbar \, \Pi(p^2) \, ,
\ee
where the left side of this relation depends implicitly on $\Pi(p^2)$. This makes the 2PI formalism 
very powerful. The self-energy $\Pi(p^2)$ is determined by a non-linear relation even if $\Gamma_2$ is truncated at a fixed loop order. This can lead to a qualitatively different 
behavior as for example a mass gap. However, in the current case this is not possible since the Goldstone mass
is protected by the Goldstone theorem.

In our discussion so far, we resummed the effects coming from heavy degrees of freedom. The resulting 
effective potential is infrared finite for dimensional reasons. The 2PI effective action on the other hand
will not only resum these effects but also the self-energy arising from interactions between light degrees of 
freedom. In particular, at one-loop the Goldstone loop is resummed [namely the first term in (\ref{pig})] but in a 
non-linear fashion. This makes the 2PI procedure more complete but also less transparent, since the IR problems 
arise only through interference with heavy degrees of freedom.

\section{A Higgs Boson Catastrophe?}
\label{HBC}

If one insists that the effective potential should be  a well-defined function of the
background field value $\phi$ not only around its minimum (the issue discussed
so far) but in general, then one should also worry about a different type of IR divergences: those
that afflict the potential when $H\rightarrow 0$, that is, at the point in field space
at which the curvature of the Higgs potential changes sign, from convex to concave.
For a Higgs mass not particularly light, like the one realized in the SM, this point
is well below the electroweak minimum, at $\phi\sim v/\sqrt{3}$, so that one can study the potential minimization
 without having to worry about it. For other applications which actually depend on the
shape of the potential, like phase transitions in the early universe or for 
inflaton potentials, this IR divergence might be dangerous and a cure for it should be
found. In this case however, it is important to keep in mind that now we are not dealing with
a physical quantity: unlike the value of the potential at the minimum, the shape of the potential
is even gauge-dependent. This means in particular that the cure for this type of divergence
might well come {\it e.g.} from radiative corrections to the kinetic term of the scalar field 
(Higgs, inflaton or otherwise).

In any case, let us examine what form these divergent terms take in the SM and explore
whether a simple resummation like the one presented in previous sections for the $G\rightarrow 0$ case
 could already
fix the problem, at least at the level of insuring an IR-safe potential and first derivative.
The one-loop CW potential (\ref{CW1}) does not give trouble in this respect. Consider
next the two-loop potential. 
Top-Higgs contributions give
\be
\delta_t V_{HL_H}^{(2)}=-N_c \kappa^2h_t^2 H L_H T (3L_T-1) \ .
\ee
Higgs-Goldstone corrections give
\be
\delta_{HG}  V_{HL_H}^{(2)}= \frac{3\kappa^2\lambda}{4}HL_H\left[2G(L_G-1)-2GL_G+3G(4-L_H)
\right]\ ,
\ee
where the first term comes from $H$-$G$ diagrams, the second from $H$-$G$-$G$ diagrams and the third from
an $H$-$H$-$H$ setting-sun diagram.
Finally, the gauge-Higgs contributions give
\be
\delta_{gH} V_{HL_H}^{(2)}= \frac{3\kappa^2}{8}HL_H\left[2g^2 W (3L_W+1)+g_Z^2 Z (3L_Z+1)
\right]\ .
\ee
To perform a resummation of these IR-problematic terms similar to the one done for $G$ terms in previous sections we need the
Higgs 2-point function. At zero external momentum this is given simply by the second derivative of the potential. At one-loop 
one gets
\bea
\Pi_H &=& -2 N_c h_t^2 T(3L_T-1) + \frac{3}{2}g^2W(3L_W+1)+\frac{3}{4}g_Z^2Z(3L_Z+1)
\nonumber\\
&+&
3\lambda\left[G(L_G-1)+H(L_H-1)+(H-G)L_G+3(H-G)L_H\right]\ .
\eea
The last line contains the scalar contributions, which we keep separate to better track their origin: the first (second) term comes from
a Goldstone (Higgs) loop with a single propagator, while the third (fourth) term comes from a Goldstone (Higgs) loop with two propagators. The prefactor $(H-G)$ in the latter two terms comes from using the relation $2\lambda\phi^2=H-G$.

\begin{figure}[t]
\begin{center}
\includegraphics[width=0.158\textwidth]{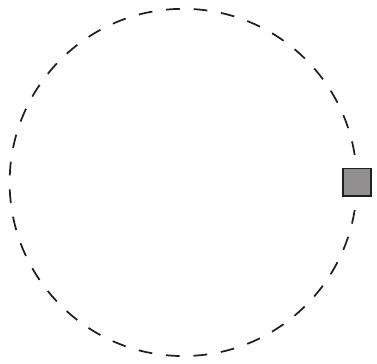}\hspace*{1cm}
\includegraphics[width=0.165\textwidth]{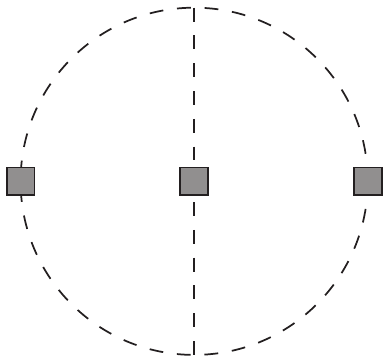}\hspace*{1cm}
\includegraphics[width=0.315\textwidth]{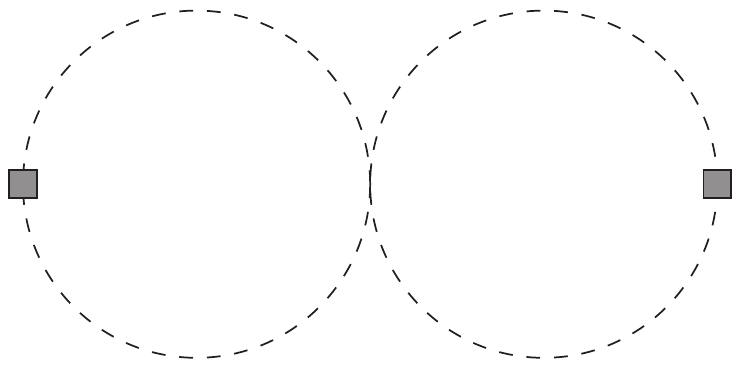}
\end{center}
\begin{center}
\parbox{14cm}{
\caption{
\label{fig:hIR}
\emph{Resummed Higgs contributions to the effective potential at one and two loops to cure $H\rightarrow 0$ IR problems. The square in the Higgs propagator
indicates a corrected Higgs propagator, as explained in the text. 
}}}
\end{center}
\end{figure}
 
It can then be checked that a one-loop resummed CW potential with the usual Higgs contribution
modified by $H\rightarrow H+\Pi_H$ will resum all the two-loop $HL_H$ terms shown above,
except the one coming from the $H$-$H$-$H$ setting-sun diagram. The reason for this mismatch is the 
usual one: the symmetry factors for two-loop diagram and Higgs 2-point function do not match.
The solution should be clear by now: use only the part of the Higgs two-point function that involves heavy fields
(in the $H\rightarrow 0$ limit), dropping the contributions from $H$ loops. That is, use
\be
V^{(1,R)}_{H}=\frac{3\kappa}{4}(H+\kappa\Pi_h)^2\left[\log\left(\frac{H+\kappa\Pi_h}{Q^2}\right)
-\frac{3}{2}\right]\ ,
\label{CW1RgoodH}
\ee 
where 
\be
\Pi_h = -2 N_c h_t^2 T(3L_T-1) + \frac{3}{2}g^2W(3L_W+1)+\frac{3}{4}g_Z^2Z(3L_Z+1)
+3\lambda\left[HL_G-G\right]\ .
\label{pih}
\ee
The two-loop $H$-$H$-$H$ setting-sun diagram is then resummed separately, evaluating it with $H\rightarrow H+\Pi_h$, see Fig.~\ref{fig:hIR}. In the language of the method of regions used in the previous section, one can show that the divergent terms in $I(H,H,H)$ come from all Higgses being soft, $I(H_s,H_s,H_s)$ while mixed terms $I(H_s,H_h,H_h)$ evaluate to zero. In principle there should be no obstacle to performing the same kind of resummation at arbitrary order in this particular case, with the difference that now the light degrees of freedom are the (soft) Higgs plus photons and gluons. Notice that the Goldstones should be considered formally  as heavy degrees of freedom, even though their squared mass is in fact negative in this field region.
Nevertheless, we expect that a resummation of Higgs IR divergent terms along similar lines as for Goldstones is feasible.

\section{Logarithmic Higgs Potentials from Broken Scale Invariance?}
\label{LogPot}

In previous sections we have encountered IR artifacts in the effective potential that went away after 
appropriate resummation. The form of the resummed potential, a momentum integral of the logarithm
of a corrected propagator that mixes light and heavy sectors, shows up in many different 
contexts. One recent example is the scenario considered in Ref.~\cite{Abel}, in which exact scale invariance
in the UV is broken in a hidden sector and transmitted via gauge mediation to the observable sector.

One of the main results of Ref.~\cite{Abel} is that the heavy UV physics can have sizable effects on the 
Higgs effective potential through a momentum integral of the logarithm of the propagator
of gauge bosons, in which IR and UV physics mix. Explicitly, the effective potential in this kind of model 
is calculated in Ref.~\cite{Abel} via the resummed expression for the gauge boson contributions as
\be
\label{VAbel}
V = \frac32 \mu^{2\epsilon}\, {\rm Tr} 
\int \frac{d^Dp_E}{(2\pi)^D} \log \left( p_E^2 + m_V^2 + \Pi_V \right)  \, ,
\ee
where we write the integration over Euclidean momentum, the trace is over gauge group indices, 
$m_V$ is the ($\phi$-dependent) gauge boson mass and
\be
\Pi_V = p_E^2 (g^2 C_{vis} + g^2 C_{hid}) \, ,
\ee
comes from one-loop radiative corrections to the gauge boson propagators, with $C_{vis}$ ($C_{hid}$) capturing 
the corrections from visible (hidden) sector loops. One basic property of these functions is that $C_{vis}$ depends on the Higgs field $\phi$ while $C_{hid}$ does not
and depends instead on the scale $f_c$ at which scale invariance is broken. 
Moreover, the hidden-sector contribution to $\Pi_V$ above should not generate
non-zero masses for the gauge bosons so that $C_{hid}$ should not have singularities at $p_E\rightarrow 0$.
This is not required of $C_{vis}$, which in principle could have $1/p_E^2$ terms, corresponding to radiative
corrections to gauge boson masses.

Using next a high momentum expansion of $C_{vis}(\phi,p_E^2)$, Ref.~\cite{Abel} obtained the Higgs effective potential induced by the gauge-mediation of scale breaking. It has the form 
\be
\label{AbelV}
 V = \frac{1}{2}a\kappa^2 f_c^2 \phi^2+ \frac{1}{2}b\kappa^3 f_c^2\phi^2\log\left(\frac{\phi^2}{f_c^2}\right)+\frac{1}{4}\lambda\phi^4\, ,
\ee
with $\kappa^2 f_c^2$ required to be of electroweak size. Here, $a$ and $b$ are two constants that depend on the details of the UV hidden sector and the gauge groups involved in gauge-mediating scale breaking.

The potential (\ref{AbelV}) deviates from the tree-level form it takes in the SM by the presence of a logarithmic-mass term and 
Ref.~\cite{Abel} identified this as a distinctive feature peculiar to this kind of scenario that would lead to non-standard implications
 for phenomenology (and also for cosmology \cite{josemi}). Notice that this logarithmic
term is of the same form of the $GL_G$ or $HL_H$ terms we have discussed in previous sections of this paper: in the conformal case $m^2=0$ so that $G=\lambda\phi^2$, $H=3\lambda\phi^2$, and $GL_G\sim HL_H\sim \phi^2\log\phi^2$ . However, there are very important differences. First, this term would cause an IR divergence only in $V''$ and only at $\phi\rightarrow 0$, rather than at the electroweak vacuum. Second, this term does not appear through IR-divergent Goldstone or Higgs loops but rather via  loops of gauge bosons. This implies that such terms cannot be resummed into IR safe terms as we did before, because gauge boson masses will always be proportional to $\phi$.   
 
Nevertheless, the appearance of this logarithmic term raises some concerns as it is in sharp contrast with the 
result expected from general principles in low-energy effective field theories (EFT). In particular, all physical effects from 
decoupling a heavy UV sector should be suppressed by the UV scale. Whenever a non-decoupling effect is found, it can only 
lead to a renormalization of the low-energy EFT parameters~\cite{Appelquist:1974tg}~\footnote{A notable exception is 
the decoupling of a fermion by increasing its Yukawa coupling. If it contributes to an axial anomaly, the decoupling can lead to additional Wess-Zumino-Witten terms in the effective action~\cite{D'Hoker:1984ph}.} and is therefore not measurable.
In fact, the threshold effects that leave an imprint of the UV physics in the low-energy physics are always analytic in the
low-energy parameters (like $m_V\sim\phi$ here) which forbids a $\phi^2\log\phi^2$ term (see {\it e.g.} Ref.~\cite{Georgi}) as 
a result of a threshold effect. In fact, logarithms of ratios of mass  scales are associated to renormalization group running 
in the effective theory between those two scales. The end result is that, after properly absorbing the heavy physics effects
in renormalizable terms of the low-energy Lagrangian, what remains are just higher order operators suppressed by the
heavy scale.

Let us illustrate this with a simple toy model (see {\it e.g.} Ref.~\cite{Burgess:2007pt} for a detailed review)
that has some of the key features of the scenarios of gauge-mediated broken scale invariance. We consider a light degree of freedom $\phi$ coupled to a heavy degree of freedom $\Phi$. The Lagrangian reads
\be
\label{lagr_toy}
{\cal{L}} = \frac12 \partial_\mu \phi \partial^\mu \phi 
+ \frac12 \partial_\mu \Phi \partial^\mu \Phi 
- \left(\frac12 m^2 \phi^2 + \frac12 M^2 \Phi^2 +\frac{1}{4}\lambda\phi^4
+ f \, \phi^2 \Phi\right)\, .
\ee
The masses of the two fields satisfy 
$m^2 \ll M^2$ and the dimensionful coupling $f$ is also heavy but does not necessarily 
become large in the decoupling limit we consider.

The low-energy EFT below the $M$ threshold is obtained either by solving the equations of motion for the heavy scalar 
or by matching the observables in the EFT to the full UV theory. The equation of motion for $\Phi$ in 
leading order (tree level and zero momentum) reads
\be
\Phi = - \frac{f}{M^2} \phi^2 \, .
\ee
Substituting this back in \eq{lagr_toy} we see that the only effect of the heavy particle in the potential for $\phi$ is to induce a shift in its quartic coupling: $\lambda\rightarrow \tilde\lambda=\lambda -2 f^2/M^2$. This coupling does not 
only reproduce the 2-by-2 $\phi$ scattering correctly, but in fact all scattering amplitudes 
in leading order. From the point of view of low-energy physics, the shift in $\lambda$ cannot be measured.

Going beyond tree-level in the matching of the effective theory to the full one, one gets at one-loop
contributions to the mass and kinetic term of the light field $\phi$ that scale as $\delta m^2 \sim \kappa f^2 $
and $\delta Z\, p^2\phi^2\sim (f^2/M^2) p^2\phi^2$. This can be obtained directly from a simple diagrammatic
calculation or, in the case of $\delta m^2$, from the one-loop Coleman-Weinberg potential, which reads
\be
V_1 (\phi) =  \frac{\kappa}{4} H^2 \left(L_H - \frac32 \right)
+ \frac{\kappa}{4} X^2 \left( L_X - \frac32 \right)\ ,
\ee
where,  for $M^2\gg m^2$, one has $H = m^2 +3\tilde\lambda \phi^2+ {\cal O}(f^2\phi^4/M^4)$ and $X = M^2+4f^2\phi^2/M^2+{\cal O}(f^2\phi^4/M^4)$. An expansion of the $X$ contribution in inverse powers of 
$M$ generates a shift in the $\phi$ mass term, $m^2\rightarrow \widetilde m^2=m^2+\delta m^2$, with
\be
\delta m^2 = \kappa \Pi_\phi= 4 \kappa f^2 (L_{M^2}-1)\ .
\ee
Such effect does not decouple with large $M$, but it is simply reabsorbed into a redefinition of the mass term.
In analogy with previous discussions in this paper we also use the notation $\Pi_\phi$ for this type of correction.

At two-loops, the effective potential of the full theory includes a contribution from a setting-sun diagram
with two $\phi$ propagators and one $\Phi$ propagator that gives a term
\be
\label{harmfultoy}
\delta V^{(2)} \sim  \kappa^2  f^2 H L_H (L_X - 1) \, .
\ee
This term, which is parametrically  of the form (\ref{AbelV}),  is another instance of a contribution that is not suppressed by inverse powers of $M$. Moreover, it seems to have an impact on the dynamics of the light field as there is no such logarithmic term in the Lagrangian that can be used to absorb it. The solution to this puzzle comes from the fact that the one-loop shifts in the renormalizable parameters $\lambda$, $m^2$ and $\phi$ wave-function renormalization should be taken into account consistently in higher order loop corrections. In particular, such corrections should be functions of $\widetilde m^2$, the shifted mass. This means that the one-loop contribution
from $\phi$ loops will take the form
\bea
V &=&  \frac{\kappa}{4} (H + \kappa \Pi_\phi)^2 \left[\log\left(\frac{H + \kappa  \Pi_\phi}{Q^2}\right) - \frac32 \right] \, .
\eea
In fact, expanding this expression to higher loop order reproduces the harmful two-loop term (\ref{harmfultoy}).
In other words, the potentially non-decoupling important effect in  \eq{harmfultoy} is absorbed into the renormalized parameters
that appear in one-loop corrections.

This toy model reproduces the key features found in the analysis \cite{Abel}. Here, instead of a momentum integral
for light gauge bosons that were sensitive to heavy physics, the two-loop contribution (\ref{harmfultoy}) can be
seen to come from the expansion of a momentum integral of a light scalar field coupled
to a heavy state of the form
\be
\label{Vtoy}
V = \frac12 \mu^{2\epsilon} \int \frac{d^Dp_E}{(2\pi)^D} \log \left( p_E^2 + H + \Pi_\phi \right)  \, ,
\ee
From the discussion above, it is clear that any contribution in $\Pi_\phi(p^2)$ to $\Pi_\phi(0)$ or $\Pi_\phi^\prime(0)$  will
only contribute to the mass or wave function renormalization of the $\phi$ field and will decouple. 

Going back to the results in Ref.~\cite{Abel}, the origin of the $\phi^2\log\phi^2$ terms in the potential (\ref{AbelV}) can be traced 
back to particular pieces in the momentum
integral (\ref{VAbel}) of the form
\be
\label{VAbelsimple} 
V = \frac32 \mu^{2\epsilon}\, {\rm Tr} \int \frac{d^Dp_E}{(2\pi)^D} \log \left[ p_E^2 + m_V^2 + \alpha_{hid}\, p_E^2 + \beta_{vis}\, m_V^2
\log(m_V^2/f_c^2) \right]  \, ,
\ee
where $\alpha_{hid}$ and $\beta_{vis}$ are momentum independent. Leaving aside possible issues with the momentum expansion of the $C_{vis}$ that leads to such terms, it should be clear from the expression above that the effect of $\alpha_{hid}$ can be absorbed
in the wave-function renormalization of gauge boson fields, while the effect of $\beta_{vis}$ will be absorbed by a shift in $m_V^2$.
To show this in the expanded expression requires a  matching of the EFT parameters to the UV theory, but this was not carried out in
Ref.~\cite{Abel}. As explained above, this can lead to spurious non-decoupling terms in the effective potential. 

\section{Conclusions}

The perturbative calculation of the Standard Model effective potential in Landau gauge encounters
problematic terms that lead to infrared divergences in the limit $G\rightarrow 0$ ~\cite{Martin3L}, where
$G$ denotes the Goldstone mass squared.  At the two-loop level one finds terms of order $G \log (G/Q^2)$,
that can give parametrically large contributions to the effective potential minimization equation $V'=0$. 
At three-loop level, terms of order $\log (G/Q^2)$ give an IR divergent contribution to the potential itself, 
and the problem becomes progressively more severe, with inverse powers of $G$ appearing at higher loop orders.

Such IR-problematic terms come from diagrams with Goldstone bosons carrying small momentum, $k^2\sim G$.
The most severe divergences arise from those diagrams with the largest possible number of Goldstone propagators carrying the same
momentum, with this number growing with higher loop order.
In this paper we have proven that these divergences are spurious and have shown explicitly that the leading divergences can be resummed in a simple way by reorganizing the perturbative expansion to take into account the effect of self-energy diagrams on the Goldstone boson propagators, with $G\rightarrow \tilde G\equiv G + \Pi_g$  where $\Pi_g$ is some well-defined contribution to the Goldstone mass 
and can be calculated perturbatively to the order required [its one-loop expression is given in \eq{pi}].

Even though this looks like an obvious solution, there is a number of technical and subtle points that require care, especially when
one tries to resum all IR-divergent terms beyond the leading ones.
We have sketched in Sec.~\ref{AWOP} how to resum also those subleading divergences by means of a reorganization of all 
mixed diagrams, that is, those with heavy and light fields, which are the only ones that lead to IR problems. 
The method follows a Wilsonian-inspired ordering principle that in practice corresponds
to the use of the method of regions \cite{regions} to isolate the offending IR divergences of any diagram.  The method can be interpreted as the integrating-out of heavy particles (or light ones carrying hard momentum, $k^2\gg G$) and allows one to reduce any mixed diagram
to a set of diagrams involving only light degrees of freedom (Goldstones, photons and gluons). In this way one also automatically  identifies which 
diagrams (with resummed Goldstone propagators) will take care of the IR divergent terms in the original mixed diagram.

In summary, after resummation,  terms polynomial in $G$ or of order $\tilde G^{2+n} \log^m (\tilde G/Q^2)$,
with $n,m\geq 0$, remain in the potential while the harmful
terms of order $G \log (G/Q^2)$, $\log (G/Q^2)$ or $1/G^n$ are absorbed by resummation.  A simple recipe for achieving
this is given in \eq{recipe}. 
A similar procedure should also work to cure another IR problem of the potential, related to $H\rightarrow 0$ divergences,
where $H$ is the (field-dependent) Higgs squared mass.
Although we have applied this method to the particular example of the Standard Model effective potential, the method used is generic and would apply to any other model in which Goldstone radiative corrections lead to IR-problematic terms in the potential.

The problem of the resummation of Goldstone IR contributions to the effective potential has to do with the influence 
of heavy degrees of freedom on the propagators of light fields (Goldstones) and similar issues arise in many different contexts.
In the last section of our paper we discuss one such case: a scenario that studies a mechanism of
gauge mediation to the visible sector of scale breaking in the UV~\cite{Abel}. By studying the impact of the
heavy physics responsible for the breaking of scale invariance on the propagators of gauge bosons Ref.~\cite{Abel}
argued that  the Higgs potential will have additional (non-standard) terms of the form
\be
\delta V(\phi)\sim
f_c^2 \, \phi^2 \, \log (\phi^2/f_c^2) \, ,
\ee
where $f_c$ parametrizes some UV scale. Such terms could in fact modify the phenomenology of electroweak symmetry breaking significantly.
We argue, by using a simple toy model for illustration, that terms of this form can indeed arise from a heavy UV sector. However, they amount to contributions of the UV sector to the self-energies of light degrees of freedom. Once a resummation and matching of the parameters in the effective theory to the full UV theory is performed, these terms are absorbed in the renormalizable parameters of the low-energy theory and all their observable effects disappear.

\section*{Acknowledgments}
We thank S. Abel and A. Mariotti for correspondence regarding Ref.~\cite{Abel} 
and M. Garny, A. Papanastasiou and M. Shaposhnikov for useful discussions. 
J.R.E. and T.K. thank CERN for hospitality and partial financial support during the early stages of this work.
J.R.E. would like to express a special thanks to the Mainz Institute for Theoretical Physics (MITP) for its hospitality 
and support during the early stages of this work.
This work has been partly supported by the Spanish Ministry MICNN under grants FPA2010-17747 and
FPA2011-25948 and by the Generalitat grant 2014-SGR-1450. The work of J.E.M. has been supported by the Spanish Ministry MECD through the FPU grant AP2010-3193.

\appendix
   \numberwithin{equation}{section}

\section{Goldstone 2-point Functions}

The one-loop 2-point functions for neutral and charged Goldstone bosons are
given by 
\bea
\Pi_0(s) &=&\lambda \left[A(H)+5A(G)-2(H-G)B(H,G)\right]-h_t^2N_c \left[s B(T,T)+2A(T)\right] \nonumber\\
&+&\frac{1}{2}g^2 \left[B_{SV}(G,W)+A_V(W)\right]+
\frac{1}{4}g_Z^2\left[B_{SV}(H,Z)+A_V(Z)\right]\ ,
\label{Pi0s}
\eea
\bea
\Pi_+(s)&=&\lambda \left[A(H)+5A(G)-2(H-G)B(H,G)\right]-h_t^2N_c \left[(s-T) B(T,0)+A(T)\right] \nonumber\\
&+&\frac{1}{4}g^2\left[B_{SV}(G,W)+B_{SV}(H,W)+2A_V(W)\right]+
\frac{1}{4}g_Z^2c_{2\theta}^2 \left[B_{SV}(G,Z)+A_V(Z)\right]\nonumber\\
&+&e^2\left[B_{SV}(G,0)+W B_{VV}(W,0)+(Z-W)B_{VV}(W,Z)\right]\ ,
\label{Pipluss}
\eea
where, in the last expression, we have set to zero the bottom quark mass and we introduced $s=p^2$.
Moreover, $g_Z^2=g^2+{g'}^2$, $e^2=g g'/g_Z$ and $\sin\theta=g'/g_Z$.  The loop functions 
we have used, following the notation in Ref.~\cite{martin2pi}, are given by:
\bea
A(X)&=&X(L_X-1)\ ,\nonumber\\
B_{SV}(X,V)&=&(2X-V+2s)B(X,V)+A(X)-A(V)\nonumber\\
&&+\frac{1}{V}\left\{(X-s)A(V)-(X-s)^2\left[B(X,V)-B(X,0)\right]\right\}\ ,\nonumber\\
A_V(V)&=&3A(V)+2V\ ,\nonumber\\
B_{VV}(X,Y)&=&\left\{-\frac{5}{2}+\frac{1}{4X Y}\left[2s(X+Y)-X^2-Y^2-s^2\right]\right\}B(X,Y)+2+
\frac{A(X)}{4X}+\frac{A(Y)}{4Y}\nonumber\\
&&+\frac{1}{4X Y}\left[(X-s)^2B(X,0)+(Y-s)^2B(0,Y)-s^2 B(0,0)\right]\ ,
\eea
where $L_X=\log(X/Q^2)$, with $Q$ the renormalization scale and we have explicitly used Landau gauge.
The function $B(X,Y)$ is
\be
B(X,Y)=-\int_0^1\log\left\{\frac{1}{Q^2}\left[X x+Y(1-x)-s x(1-x)\right]\right\}\, dx\ , 
\ee
and $s$ is assumed to have a small positive imaginary part. An explicit analytical expression
for $B(X,Y)$ is well known and can be found {\it e.g.} in Ref.~\cite{Bxy}.  
It is also useful to give the explicit expressions of some of the previous functions 
in cases with some massless argument, like
\bea
B(X,0)&=&2-L_X+\left(\frac{X}{s}-1\right)\log\left(1-\frac{s}{X}\right)\ ,\nonumber\\
B(0,0)&=&2-\log\left(\frac{-s}{Q^2}\right)\ ,\nonumber\\
B_{SV}(X,0)&=&3(X+s)B(X,0)+3A(X)-2s\ ,\nonumber\\
B_{VV}(V,0)&=&\frac{3}{4V}\left[(s-3V)B(V,0)-s B(0,0)\right]+2\ .
\eea

The low-momentum expansion of the Goldstone 2-point functions in \eq{Pi0s} and (\ref{Pipluss}) reads
\be
\Pi_0(s) =  \Pi_G + s \Pi'_0 +{\cal O}(s^2)\ ,\quad
\Pi_+(s) =  \Pi_G + s \Pi'_+ +{\cal O}(s^2)\ ,
\ee
with $s=p^2$. The zero momentum part is 
\be
\Pi_G = 3 \lambda \left[G (L_G-1)+H(L_H-1)\right] -2N_c h_t^2 T(L_T-1)
+\frac{3g^2 }{2} W(L_W-1/3)+\frac{3g_Z^2}{4}  Z (L_Z-1/3)\ ,
\ee
while the next terms are:
\bea
\Pi'_+&=&\frac{\lambda}{(H-G)^2}\left[G^2-H^2+2HG(L_H-L_G)\right]
+h_t^2N_c \left(L_T-\frac{1}{2}\right)
\nonumber\\
&+&\frac{3}{4} g^2\left(\frac{5}{3}-\frac{WL_W-G L_G}{W-G}-\frac{WL_W-H L_H}{W-H}\right)+
\frac{3}{4} g_Z^2\left(\frac{5}{6}-\frac{ZL_Z-G L_G}{Z-G}\right)\nonumber\\
&+&\frac{3}{4} e^2\left\{
L_W-L_Z+2\frac{\left[W^2-Z^2-2W Z(L_W-L_Z)\right]}{(Z-W)^2}+4\frac{Z L_Z-G L_G}{Z-G}-\frac{10}{3}
\right\}\nonumber\\
&+&\frac{1}{2}e^2(5-6L_G)+\frac{1}{8}e^2\left[-15+6\log\left(\frac{-s}{Q^2}\right)-6L_W\right]\nonumber\\
&\simeq &-\lambda+h_t^2N_c \left(L_T-\frac{1}{2}\right)
+\frac{3}{4} g^2\left(\frac{5}{3}-L_W-\frac{WL_W-H L_H}{W-H}\right)+
\frac{3}{4} g_Z^2\left(\frac{5}{6}-L_Z\right)\nonumber\\
&+&\frac{3}{4} e^2\left\{
L_W+2\frac{\left[W^2-Z^2-2W Z(L_W-L_Z)\right]}{(Z-W)^2}+3L_Z-\frac{10}{3}
\right\}\nonumber\\
&+&\frac{1}{2}e^2(5-6L_G)+\frac{1}{8}e^2\left[-15+6\log\left(\frac{-s}{Q^2}\right)-6L_W\right]\ ,
\label{Pipplus}
\eea
\bea
\Pi'_0&=&\frac{\lambda}{(H-G)^2}\left[G^2-H^2+2HG(L_H-L_G)\right]+h_t^2N_c L_T
\nonumber\\
&+&\frac{1}{4}g^2 \left(5-6\frac{WL_W-G L_G}{W-G}\right)+
\frac{1}{8}g_Z^2 \left(5-6\frac{ZL_Z-H L_H}{Z-H}\right)\nonumber\\
&\simeq &-\lambda+h_t^2N_c L_T
+\frac{1}{4}g^2 (5-6L_W)+
\frac{1}{8}g_Z^2 (5-6L_Z)\ ,
\label{Pip0}
\eea
where the final approximate expressions have been expanded up to ${\cal O}(G^0)$. In $\Pi'_+$, we have
explicitly separated the last two terms which correspond to the contributions from a $G^+$-$\gamma$ loop
(next-to-last term) and a $W^+$-$\gamma$ loop (last term) to the 2-point function.


\end{document}